\documentclass[]{spie}  
\pdfoutput=1
 
\usepackage{amsmath,amsfonts,amssymb}
\usepackage{graphicx}
\usepackage[colorlinks=true, allcolors=blue]{hyperref}

\title{A simple optimized amplitude pupil mask for attempting to direct imaging of Proxima b with SPHERE/ZIMPOL at VLT
}

\author[a]{Polychronis Patapis}
\author[a]{Jonas K\"uhn}
\author[a]{Hans Martin Schmid}
\affil[a]{Institute of Particle Physics and Astrophysics, ETH Zurich, Wolfgang-Pauli Str. 27 8093, Switzerland}

\authorinfo{Further author information: (Send correspondence to P.P.)\\
E-mail: polychronis.patapis@phys.ethz.ch, Telephone: +41 44 633 68 35}

\begin{document} 
\maketitle

\begin{abstract}
Proxima b is a terrestrial exoplanet orbiting in the habitable zone of our closest star Proxima Centauri. The separation between the planet and the star is about 40 mas and this is with current instruments only reachable with direct imaging, using a visual extreme AO system like SPHERE/ZIMPOL. Unfortunately, the planet falls under the first airy ring at 2$\lambda$/D in the I band, which degrades achievable contrast. We present the design, optical simulations and testing of an amplitude pupil mask for ZIMPOL that reshapes the PSF, increasing the contrast at $r = 2\lambda$/D about an order of magnitude. The  simple mask can be inserted directly into the current setup of SPHERE.

\end{abstract}


\section{INTRODUCTION}
\label{sec:intro}  

The last decade has sparked a revolution in the field of exoplanet science with technologies and observing and data processing techniques that have led to the discovery of over three thousand planets. It is now suspected that at least every second star has one or more planets, and the diversity of the planet parameters of radii, mass and system architecture is astonishing. Several exo-earth candidates have been discovered, terrestrial planets which are located in the habitable zone of the star, have atmospheres and could have liquid water on their surface. None of these objects has been directly observed but rather found indirectly through transit and radial velocity methods. Direct imaging of exoplanets has been restricted to hot gas giants with masses a few times the mass of Jupiter. This is mainly due to the extremely challenging task of detecting faint signals, many orders of magnitude smaller than the star, which favors the discovery of self luminous young giant planets far away from the star.

A recently discovered terrestrial planet of great interest is Proxima Centauri b (HIP
70890 b) \cite{Anglada-Escude2016}, which was discovered through radial velocity method. It is orbiting around the nearest star at a distance of 1.295 parsecs with an orbit of 11.2 days and a semi-major axis of 0.048 AU. At quadrature phase the angular distance is 37 milli-arcseconds (mas) and when observed with a 8 m-class telescope it corresponds to two diffracted beam widths (2 $\lambda$/D) in the I-band. Since the host star is a red dwarf Proxima b might have conditions that make it habitable \cite{Dong2017} \cite{Meadows2016}.

\begin{figure}
\begin{center}
\includegraphics[scale=0.5]{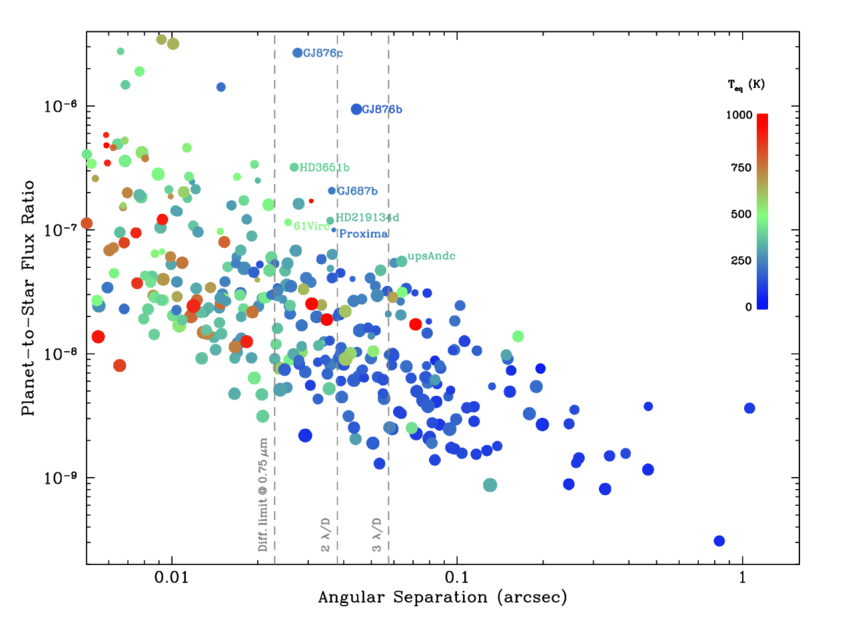}
\end{center}
\caption{Expected planet to star flux ratio in reflected light versus angular separation. The color indicates the estimated temperature of the objects and the vertical lines correspond to integer multiples of the resolving power of an 8-m class telescope at 750 nm. A few of the interesting targets are indicated by their name, including Proxima b. Plot used with permission from Lovis et al. (2016) \cite{Lovis2016}.}
\label{exoplanets}
\end{figure}

In Figure \ref{exoplanets} of Lovis et al. (2016)\cite{Lovis2016} known extrasolar planets with small separations are plotted with the y-axis indicating the expected planet-to-star flux ratio. There are various limitations that prevent current instruments to detect these objects. First, diffraction limits the wavelength range due to the current aperture size of the largest telescopes and future Extremely Large Telescopes (ELTs) that would alleviate this limit are projected in 7 to 10 years. Secondly, the requirement in contrast is orders of magnitude smaller than what is currently achieved even with the "planet-hunter" instruments that operate mainly in the Near-Infrared (H-band) and are equipped with extreme adaptive optics and coronagraphs. 

Lovis et al. (2016) \cite{Lovis2016} have proposed a coupling of the ESSPRESSO spectrograph at VLT with the high contrast imaging path of SPHERE as a High Contrast High Resolution Spectroscopy configuration and observe Proxima b with the hope to extract a spectrum of its atmosphere. As the authors suggest in this paper it would be interesting to first observe it with ZIMPOL (Zurich IMaging POLarimeter) in polarimetry on SPHERE.

\subsection*{SPHERE/ZIMPOL instrument at VLT}

ZIMPOL is a high contrast polarimeter designed to search for exoplanets in reflected light and is part of the SPHERE instrument\cite{schmid2018} at VLT. Operating in the visible to near-infrared wavelengths, with an extreme adaptive optics system \cite{fusco2014final} and using polarimetric imaging one can reach deep contrasts of up to $10^{-6}$ at close separations, whilst providing high angular resolution. Light of the star reflected off the planet's atmosphere results in a fraction of the light being polarized. ZIMPOL observes two orthogonal polarization directions simultaneously, and while the signal of the star can be found in both directions the polarized planet signal only ends up in one. One can therefore subtract the star signal efficiently with polarimetric differential imaging (PDI) and faint sources of scattered or reflected light can be detected. 

This technique has been highly successful in detecting and characterizing faint circumstellar disks\cite{engler2017hip}. An ongoing search of planets in reflected light with ZIMPOL provides for nearby bright stars planet contrast limits if $10^{-6}$ to $10^{-8}$ outside of 0.2" separation (Hunziker et al, in preparation). An important distinction of the planets searched in reflected light is that objects close to their star are favored since they receive much more irradiation, which in turn leads to stronger emission of polarized light. 

For observing Proxima b the requirements are still at the upper limit of what is theoretically achievable with the current installment of ZIMPOL. The two biggest limitations for detecting signals at separations below 3 $\lambda$/D are some uncorrected instrumental polarization induced due to the beam shift effect as described in Schmid et al (2018)\cite{schmid2018}, and the residual light from the first airy ring. The first airy ring for an obstructed aperture such as the VLT still encompasses a non-negligible fraction of the host star PSF which prevents to reach deep contrasts. At the moment the achievable polarimetric contrast is 3 $10^{-7}$ at a separation of 0.2" and the goal of this study is to provide a simple solution that can be inserted into the setup of SPHERE and would provide a similar contrast at closer separation. The idea is to design an binary amplitude pupil mask that reshapes the pupil and improves contrast in the region of interest.

\section{Optimized Pupil Apodization}

\subsection{Requirements}

The goal of this study is to provide a simple solution for suppressing the first airy ring of the VLT PSF, that provides a robust improvement to the polarimetric contrast for Proxima b in the I band. Robustness is in this case defined with respect to wavefront errors, bandwidth, effect on the polarization of the incoming beam and throughput. Therefore the following requirements were formulated:
\begin{itemize}
\item \textbf{Contrast improvement}: An improvement of contrast at the location of the exoplanet of a factor 5 to 10 was deemed realistic for a simple mask design. Also a width of 1 $\lambda$/D for the region of interest was set in order to account for the uncertainty of the expected planetary mass companion location and aberration effects that tend to push additional light to close separations to the star.
\item \textbf{Tip tilt jitter}:  Dominant aberrations are tip and tilt errors, which a pupil plane mask should be less sensitive to.
\item \textbf{Bandwidth}: The mask should be able to perform well for the different filter bandwidths of the ZIMPOL instrument, especially since a relatively broad filter (20\%) would be chosen in order to accumulate enough photons from the planet.
\item \textbf{Polarization effects}: A binary amplitude mask was chosen in order to minimize the effect of semi-transparent materials to the polarization of the incoming beam. This is important since the deep contrast achieved by ZIMPOL relies on having as little instrumental polarization introduced as possible. For the same reasons any spiders that will support the optimized mask will be placed at $45^{\circ}$ angle with respect to the measured $I_{\parallel}$ and $I_{\perp}$ polarization directions. 
\item \textbf{Throughput}: The throughput loss due to the mask should be minimized, since the polarized flux of the planet is small. A good throughput is required to achieve a good signal to noise (SNR) detection without having to integrate for a very long time\footnote{Ideally a detection should be achievable in one night.}.
\end{itemize}

\subsection{Simulation setup}
The light propagation was simulated using Fast Fourier Transforms (FFT) in Python on an oversampled grid of 4096 x 4096 and a sampling of 10 pixels per $\lambda$/D on the focal plane. The VLT aperture geometry was used as the reference PSF and all contrast and throughput calculations are normalized with respect to the maximum value of the VLT PSF core.

In order to simulate the wavefront aberrations and wavefront jitter, the phase of the incoming planar wave is multiplied with a complex term consisting of the aberrations decomposed in Zernike polynomials and a random noise distribution over the whole plane to simulate the residual RMS wavefront jitter, as a function of the Strehl ratio.

\subsection{Modified Gerchberg-Saxton algorithm}

We use a modified version of the Gerchberg-Saxton (GS) algorithm \cite{gerchberg1971phase} similar to our work in Kuhn et al. (2017)\cite{kuhn2017implementing}, but this time adjusted for optimization of the pupil plane. The algorithm works as following:
\begin{enumerate}

\item Begin with the entrance pupil given as a complex array.
\item Apply FFT to the entrance pupil to obtain the the PSF.
\item Set flux in the region of interest in the focal plane to 0.
\item Inverse FFT of the focal plane and obtain the updated complex entrance pupil
\item Take the absolute of the complex array.
\item Subtract the initial pupil from it and multiply it add 1 and multiply the entrance pupil.
\item Repeat step 2 onwards until the desired null depth in the focal plane is reached.
\end{enumerate}
\begin{figure}[ht]
\begin{center}
\includegraphics[scale=.7]{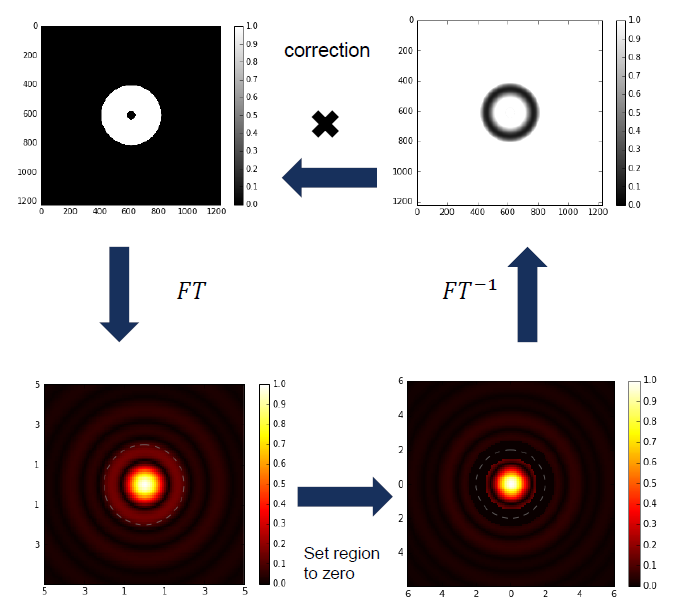}
\end{center}
\caption{Iteration of GS algorithm modified for pupil plane mask optimization.}
\end{figure}

We want to note that there are other algorithms to calculate shaped pupil masks, for example as described by Carlotti et al. (2011)\cite{carlotti2011optimal}. Due to the author's former familiarity with the used GS algorithm we chose to start experimenting with that, but we hope to also test other designs in order to find the best solution for the given problem.

For the case of Proxima b the region of interest was set as a ring in the focal plane from 1.5 $\lambda$/D to 2.5 $\lambda$/D. The number of iterations for the GS code was set to 400 and two different entrance pupils where optimized. First, the easiest scenario was the VLT pupil not taking into account its supporting structure of the secondary mirror (spiders). This is justified due to the small width of the spiders relative to the aperture radius and the fact that in first order we are interested in suppressing just the first airy ring. The second entrance pupil run with the GS algorithm included two spiders at a $45^{\circ}$ angle with respect to the horizontal axis and a width of about 4\% of the aperture radius that would provide the support for the inner mask features. This was decided after considering manufacturing limitations for the spider width, which in turn impact the performance of the original correction.

\begin{figure}[!ht]
\begin{center}
\begin{tabular}{l l}
\includegraphics[scale=.1]{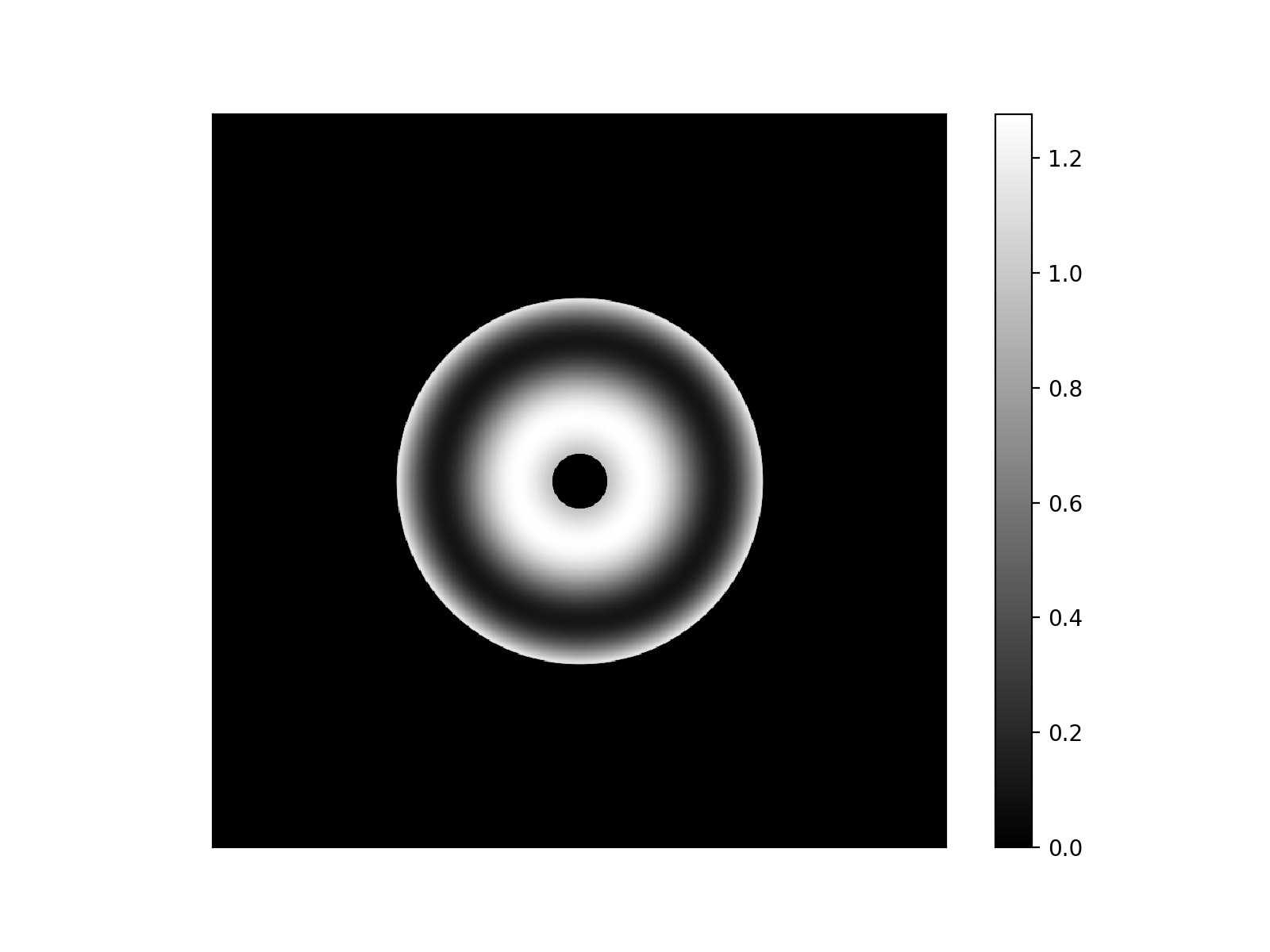}&
\includegraphics[scale=.1]{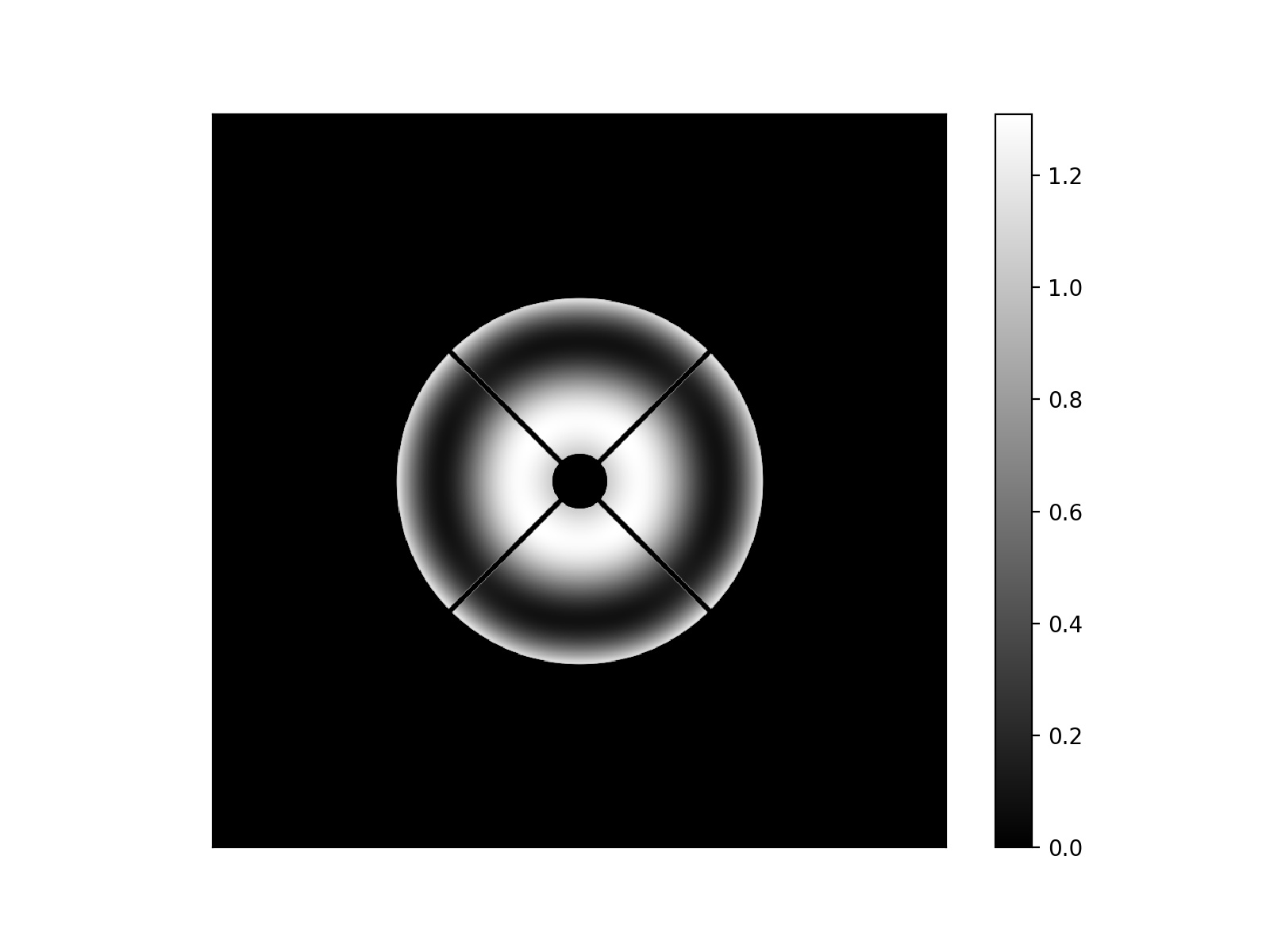}
\end{tabular}
\end{center}
\caption{Optimized mask in gray-value scale for the VLT aperture without spiders (left) and over sized diagonal spiders (right).}
\label{Grayval-masks}
\end{figure}
The output of both optimizations are shown in Figure \ref{Grayval-masks} in gray scale. The cutoff value in the gray scale for creating a binary amplitude mask (eq. \ref{cutoff}) is explored in section \ref{subsec: param-space}. The tendency of the algorithm is to produce a ring in the pupil plane that blocks the spatial frequencies that contribute to the light diffracted to the first airy ring.

\begin{align}
MASK_{binary} = 
\begin{cases}
0, \,\ MASK_{grayscale} \leq C \\
1, \,\ otherwise
\end{cases}
\label{cutoff}
\end{align}
where C is the cutoff value for the gray scale.

\subsection{Optimized mask and parameter space }\label{subsec: param-space}
The parameter space of the cutoff value C can be explored in order to examine the trade off between contrast gain over the region of interest and the throughput loss due to the insertion of the amplitude mask in the pupil plane. A contrast level of $10^{-3}$ was chosen as the goal for an unaberrated PSF in order to have some margins for performance losses due to wavefront errors, structural effects and bandwidth, while still achieving the desired improvement under real telescope conditions. A cutoff value of C=0.33 for the gray scale was chosen as an optimal value, which corresponds to a width of the ring in the pupil plane of 20.5\% of the aperture radius. For the next part of the performance assessment this value will be kept constant in order to better compare the effects of non ideal conditions. 

\section{Theoretical performance of optimized pupil masks}

\subsection{Monochromatic source}
\begin{figure}[!ht]
\begin{center}
\begin{tabular}{l l}
\includegraphics[scale=.3]{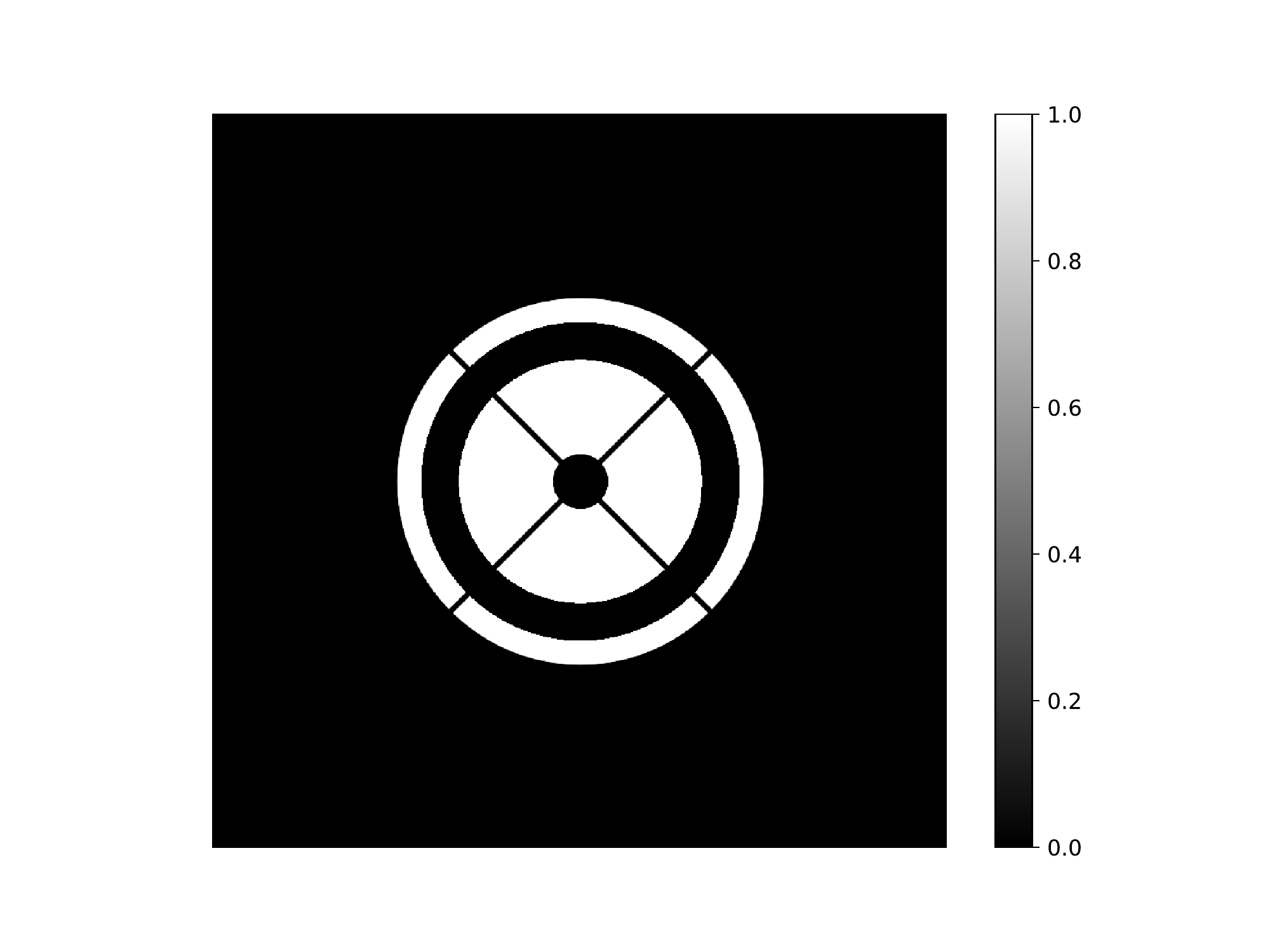}&
\includegraphics[scale=.3]{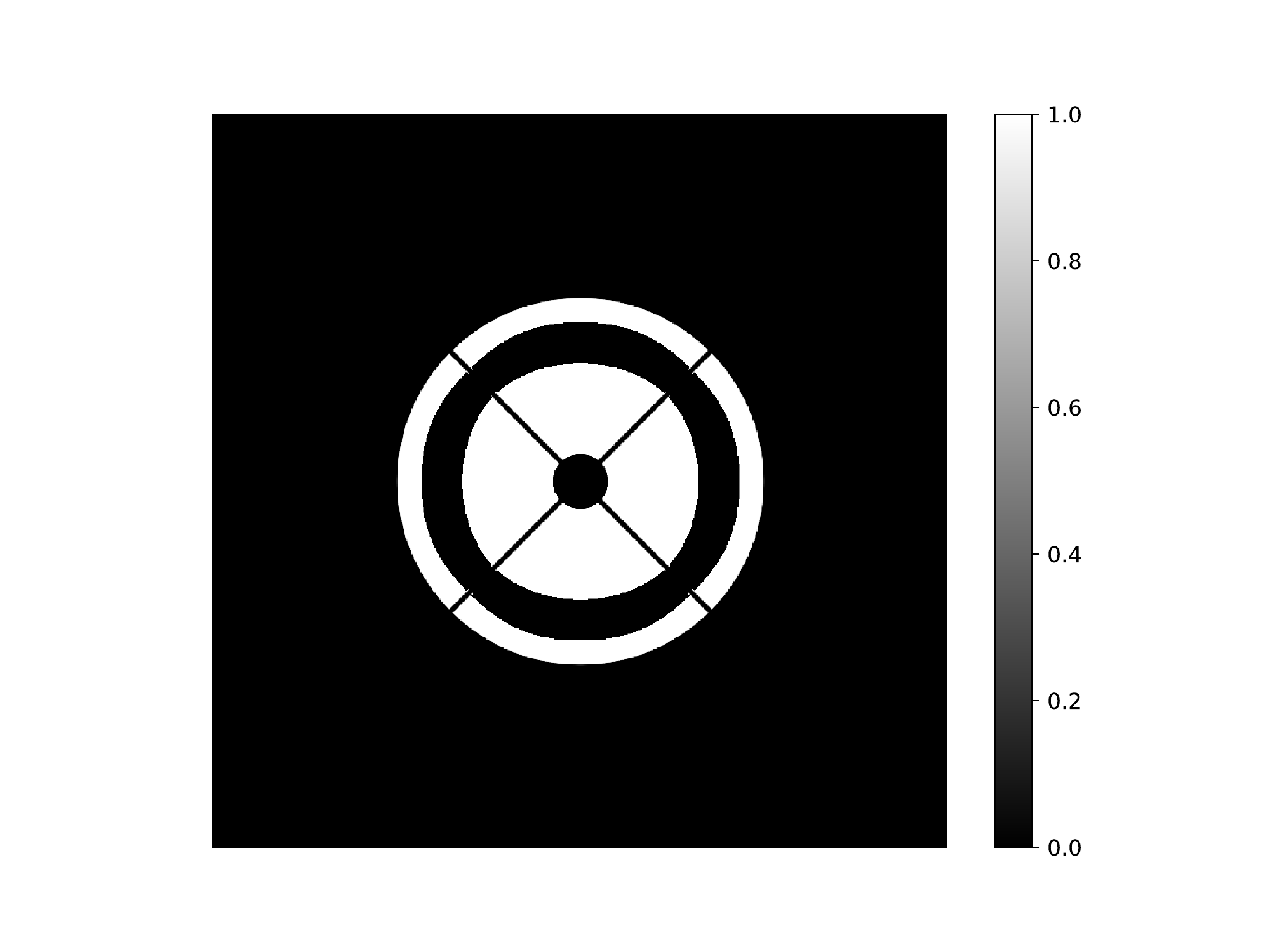} \\
\includegraphics[scale=.3]{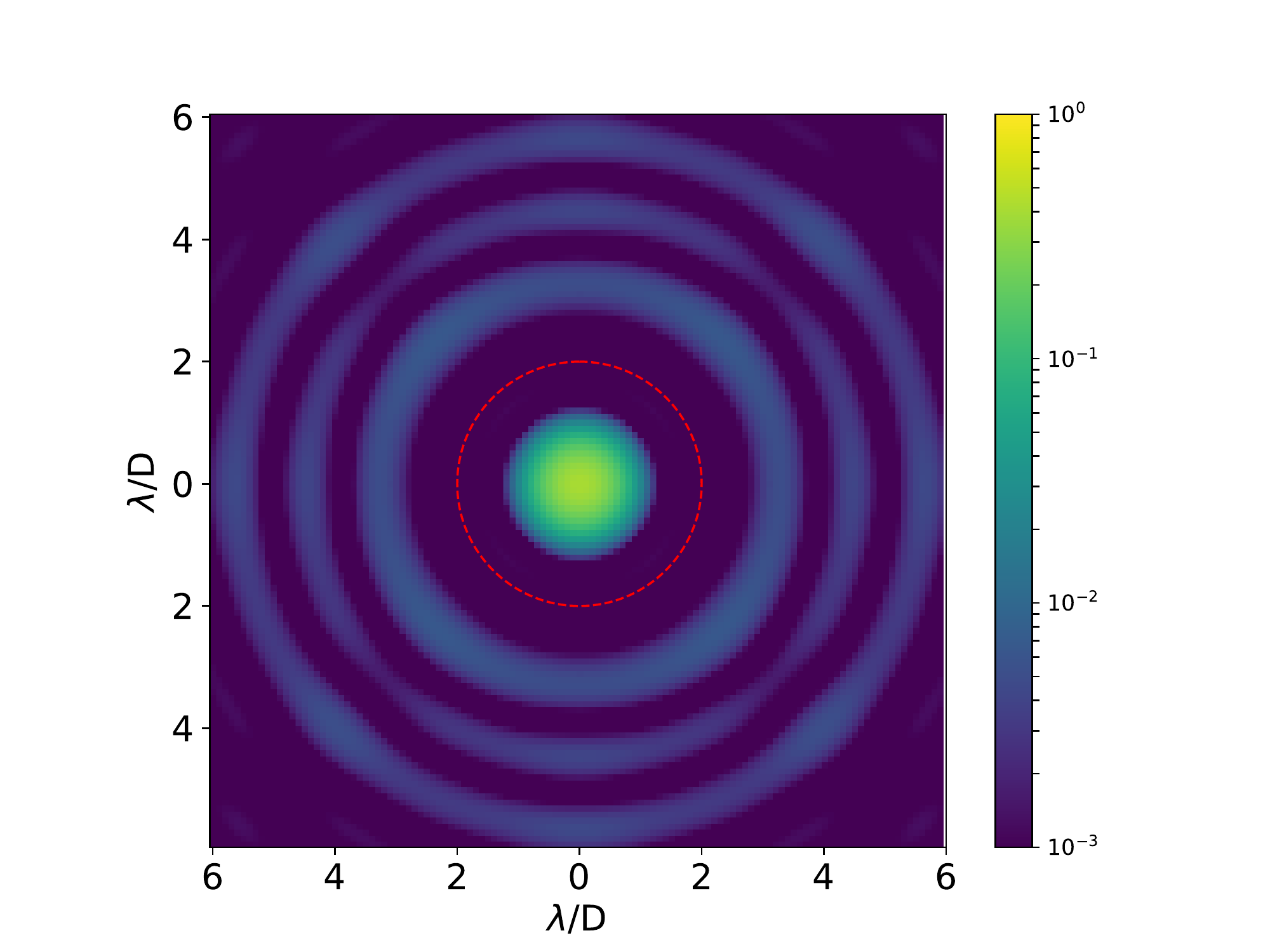} &
\includegraphics[scale=.3]{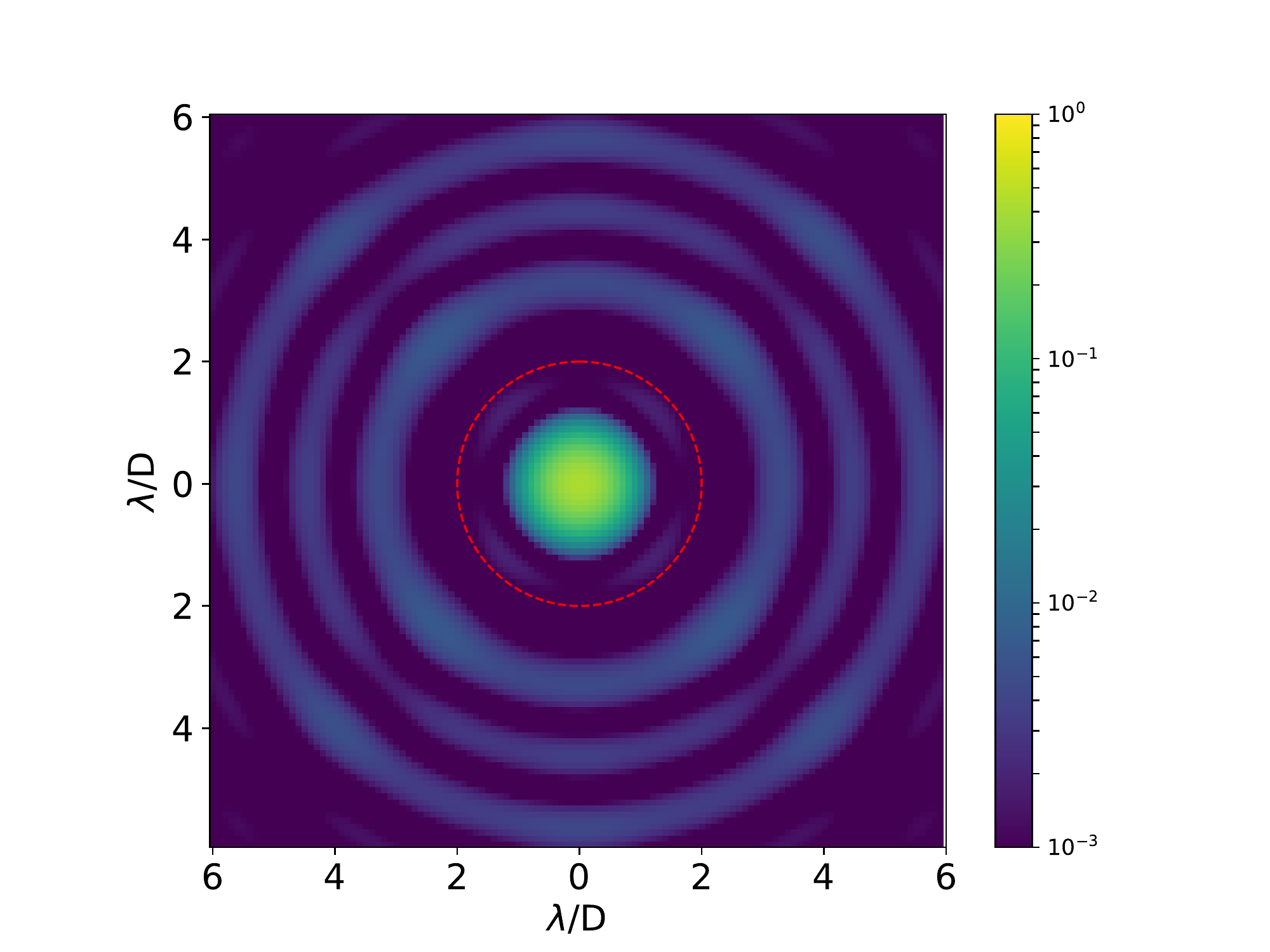}
\\
\multicolumn{2}{c}{\includegraphics[scale=.2]{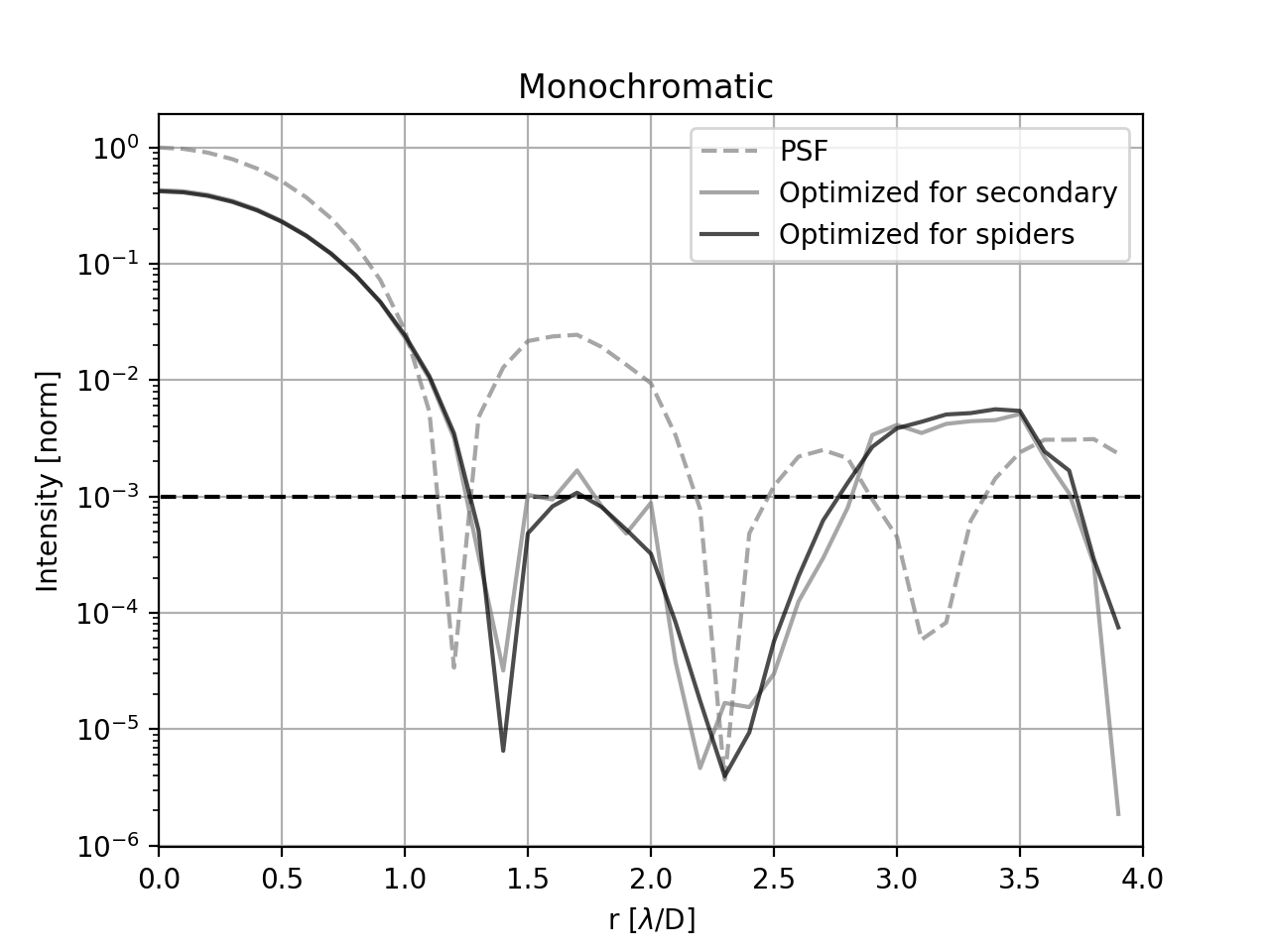}}
\end{tabular}
\end{center}
\caption{Top: Optimized binary amplitude masks without (left) and with spiders optimization (right). Bottom: PSF of the optimized mask for a monochromatic light source (750 nm) and ideal wavefront. The ref dashed circle indicates the separation of 2 $\lambda$/D}
\label{Binary-masks}
\end{figure}

The results of the ideal case of a monochromatic plane wave, for which the optimization algorithm was designed, are shown in Figure \ref{Binary-masks}. In the upper panels the binary pupil masks are shown and corresponding PSF in the middle panels. On the bottom row of Figure \ref{Binary-masks} the raw contrast curve is plotted with the nominal PSF as a photometry reference. The masks perform as expected with the region of interest being suppressed down to a contrast level of $10^{-3}$. 
\subsection{Bandwidth, Strehl ratio and aberrations}

\begin{figure}[!ht]
\begin{center}
\begin{tabular}{c c}
\includegraphics[scale=0.4]{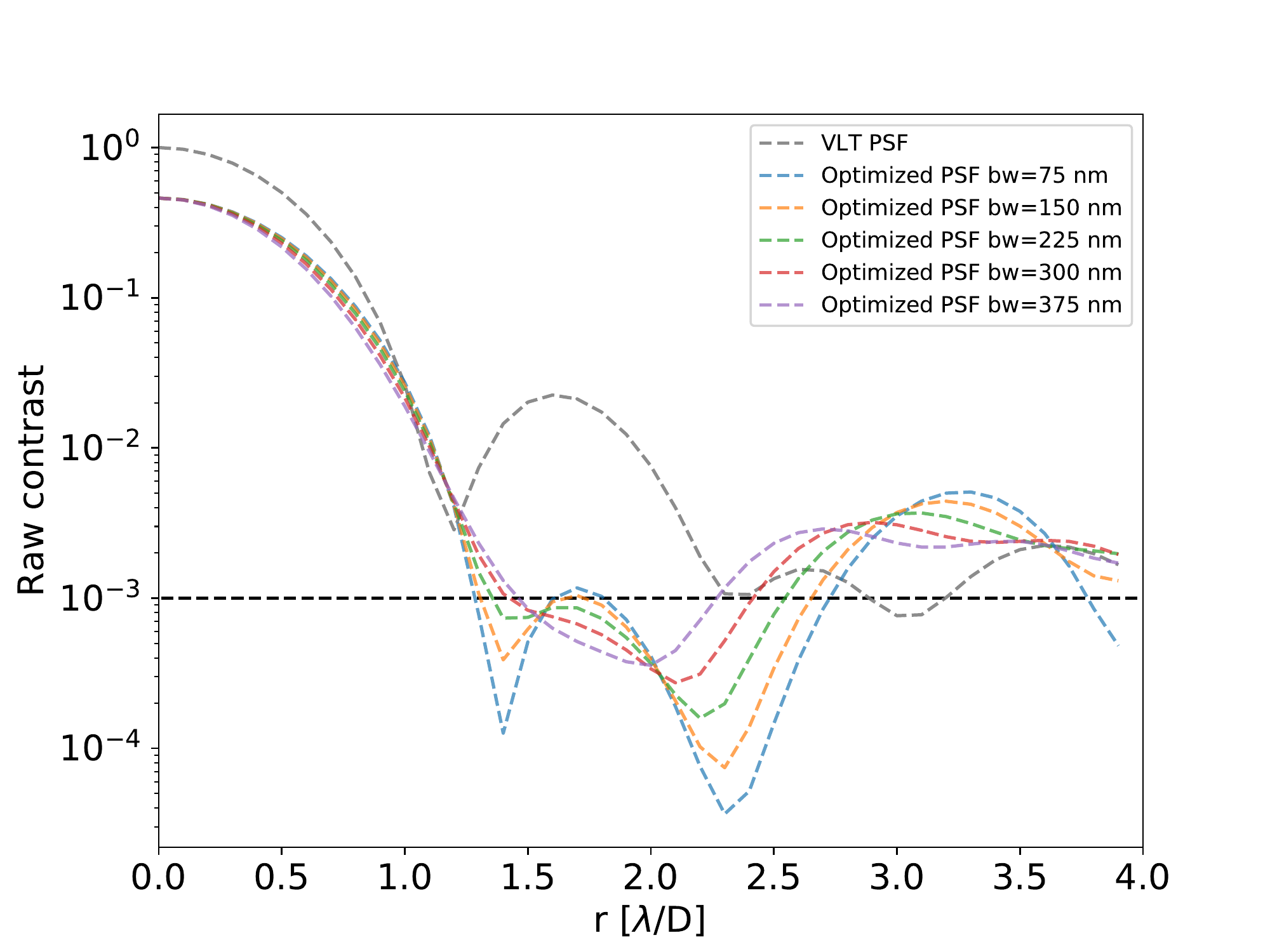} &

\includegraphics[scale=0.4]{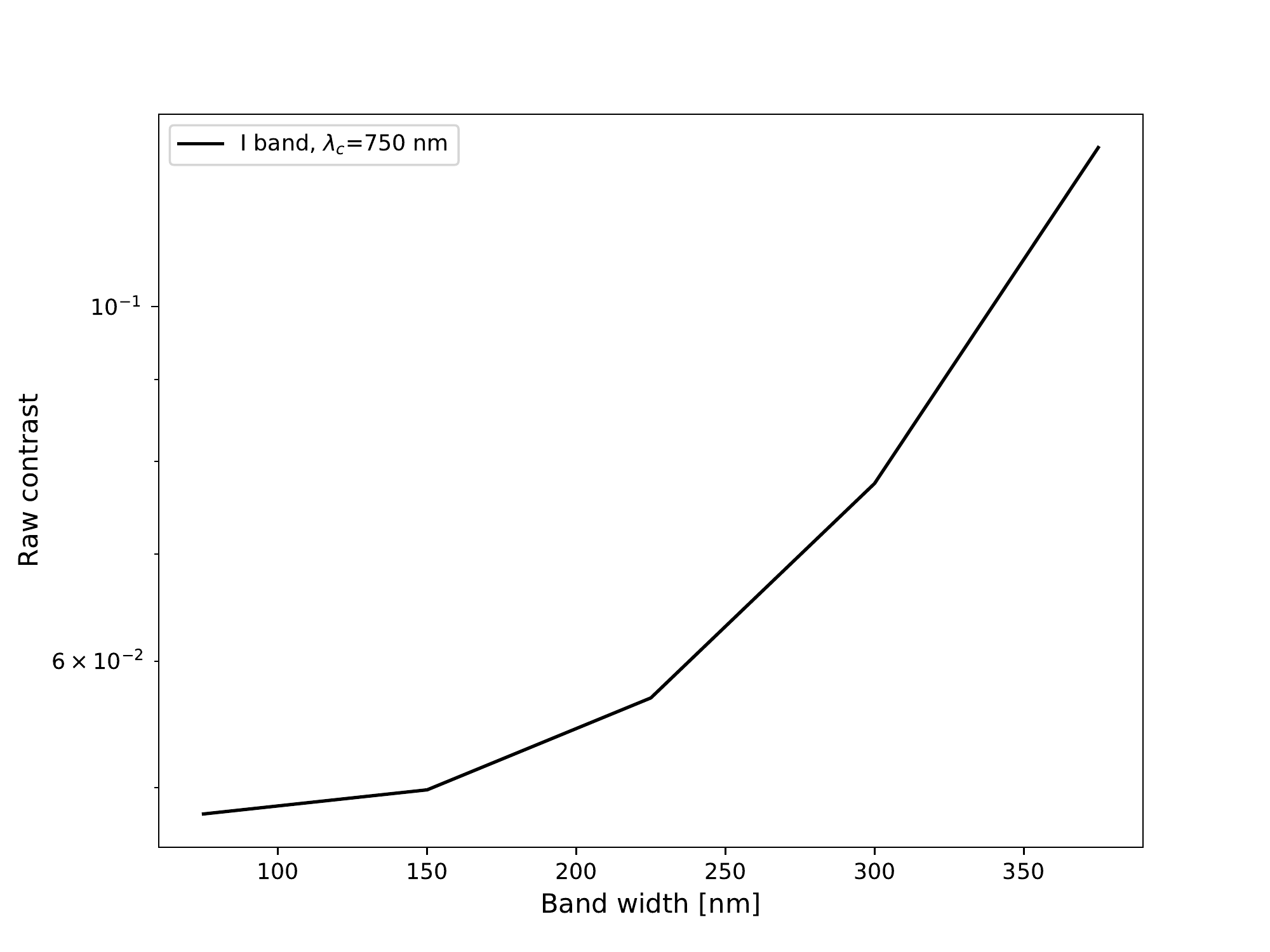} \\

\end{tabular}
\end{center}
\caption{Left: Raw contrast as a function of separation for different bandwidths centered around 750 nm. Right: Contrast in ROI as a function of bandwidth. The relevant bandwidth would be 150 nm for example for the I prime filter on SPHERE (\url{https://www.eso.org/sci/facilities/paranal/instruments/sphere/inst/filters.html})}
\label{Broadband}
\end{figure}

\begin{figure}[!ht]
\begin{center}
\begin{tabular}{c c}
\includegraphics[scale=0.4]{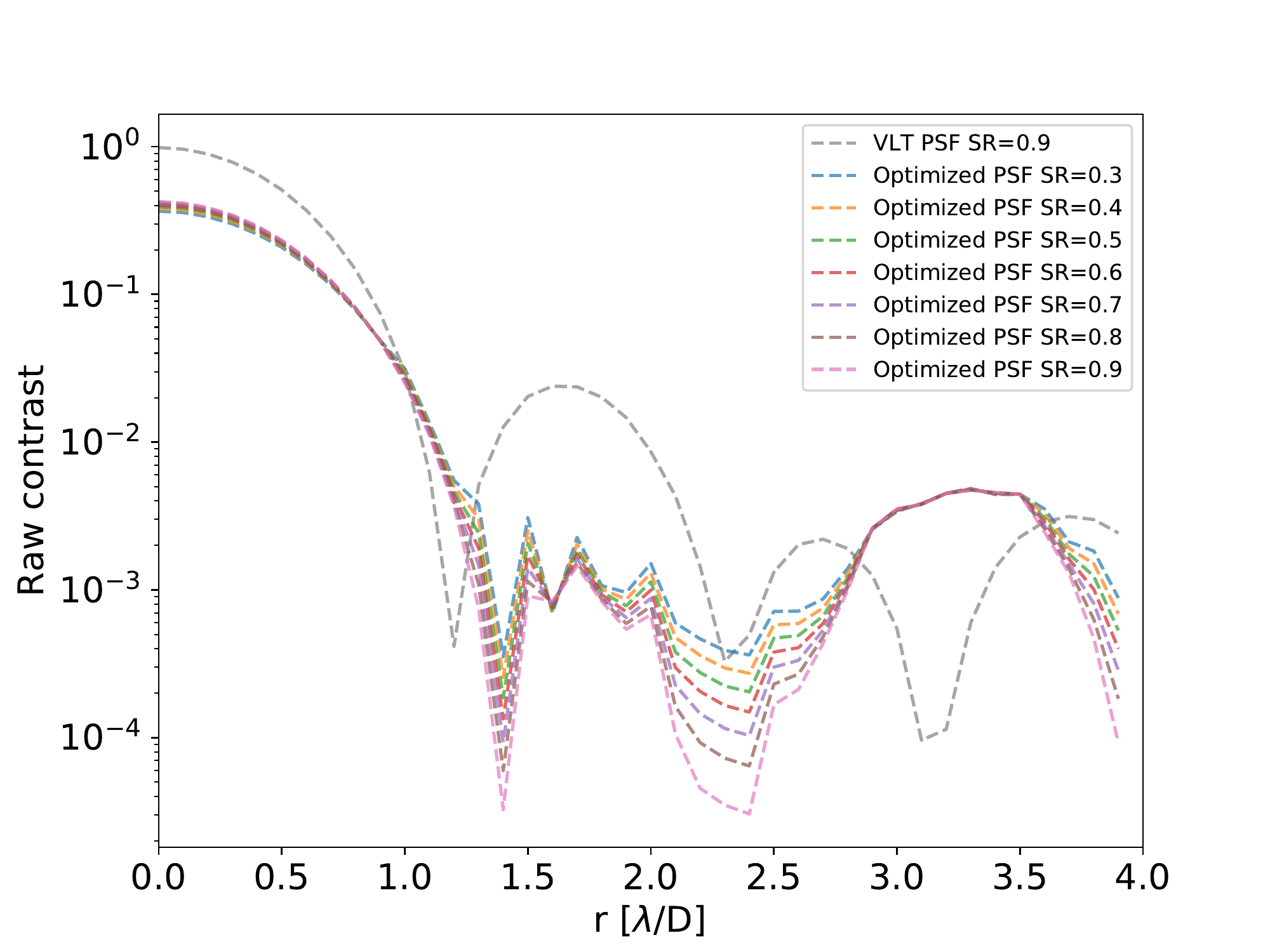} 
& \includegraphics[scale=0.4]{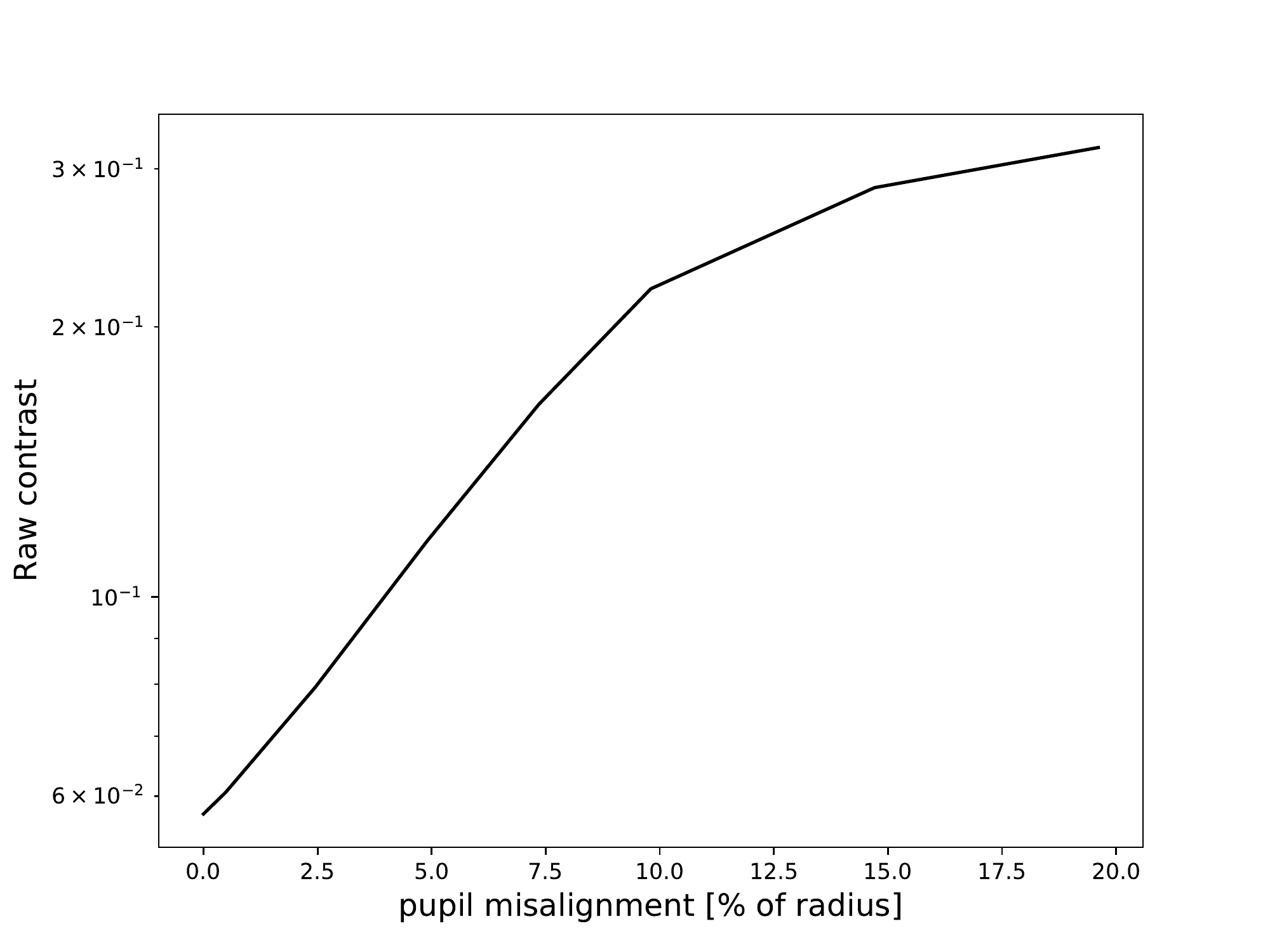}
\end{tabular}
\end{center}
\caption{Left: Raw contrast plotted against separation as a function of Strehl ratio. The distribution of noise was computed and kept constant with the Strehl ratio only affecting the magnitude of the phase error RMS of the input field. Right: Pupil misalignment impact on the contrast in the ROI.}
\label{contrast_strehl}
\end{figure}

The performance of the mask as a function of bandwidth is an important factor to consider. Although a binary amplitude mask is inherently achromatic, using it over a broad range of wavelengths will impact the contrast in the focal plane. The raw contrast is plotted against separation for different bandwidths centered around 750 nm with a perfect flat filter transmission assumed as shown in Figure \ref{Broadband}, as well as the raw contrast in the ROI as a function of bandwidth. In Figure \ref{contrast_strehl} the PSF mean contrast is plotted for different Strehl ratio is plotted to illustrate the effect of non ideal wavefront on the performance of the mask. Another effect to consider is the misalignment of the inserted mask with respect to the VLT aperture, shown in Figure \ref{contrast_strehl}.

\section{Experimental Verification}

\subsection{Optical bench}
For the experimental verification the high contrast imaging bench at ETH Zurich was used. It consists of a simple setup with a fiber coupled diode laser as a monochromatic source (633 nm), a lens to collimate the light onto the entrance pupil plane (EP), two F/250 lenses that form an intermediate focal plane and a pupil Lyot stop plane allowing to use two masks in sequence. After the second pupil plane two lenses can be inserted to image either the Lyot or the focal/science (F/400) plane. The low readout noise CCD camera\footnote{PCO Pixelfly camera: \url{https://www.pco.de/specialized-cameras/pcopixelfly-usb/}} provides for an aperture size of 5 mm a sampling of the PSF with 5 pixels per $\lambda$/D. 

\subsection{Mask manufacturing}

The masks were manufactured at the mechanical workshop of the Department of Physics at ETH Zurich, using a laser cutting machine with a custom built laser. This enables manufacturing masks with very thin features while maintaining high precision (1 $\mu$m). The masks are made from molybdenum metal type that is performing exceptionally well with this laser and has been tested for various applications. The limiting factor are the width of the spiders in the VLT and optimized pupils and masks with a 5 mm and 2.5 mm  diameter were successfully produced with different spider thickness. For testing on the bench the 5 mm masks were chosen with a spider width of 50 $\mu$m, shown in Figure \ref{Measured-PSF}. 

We note that the final design of the mask for ZIMPOL would have a 6 mm diameter and therefore this manufacturing exercise is a good proof of principle for the production process. Moreover, manufacturing more complicated designs, such as the mask optimized to deal with the spiders was successful but with a 10 mm diameter and have not yet been made and tested for a 5 mm aperture.

\subsection{Experiment results}

\begin{figure}[!ht]
\begin{center}
\begin{tabular}{l l}
\includegraphics[scale=.35]{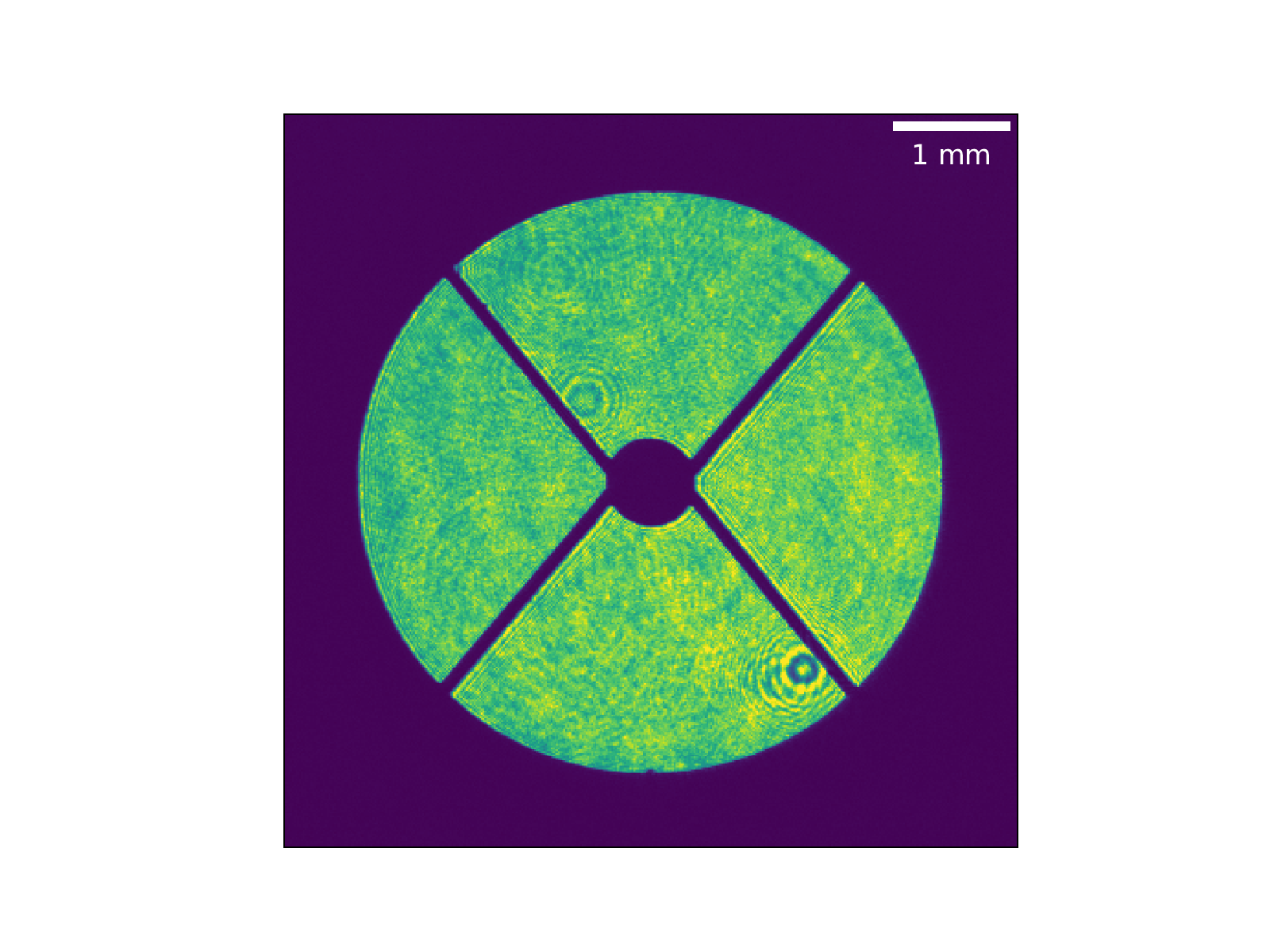}&
\includegraphics[scale=.35]{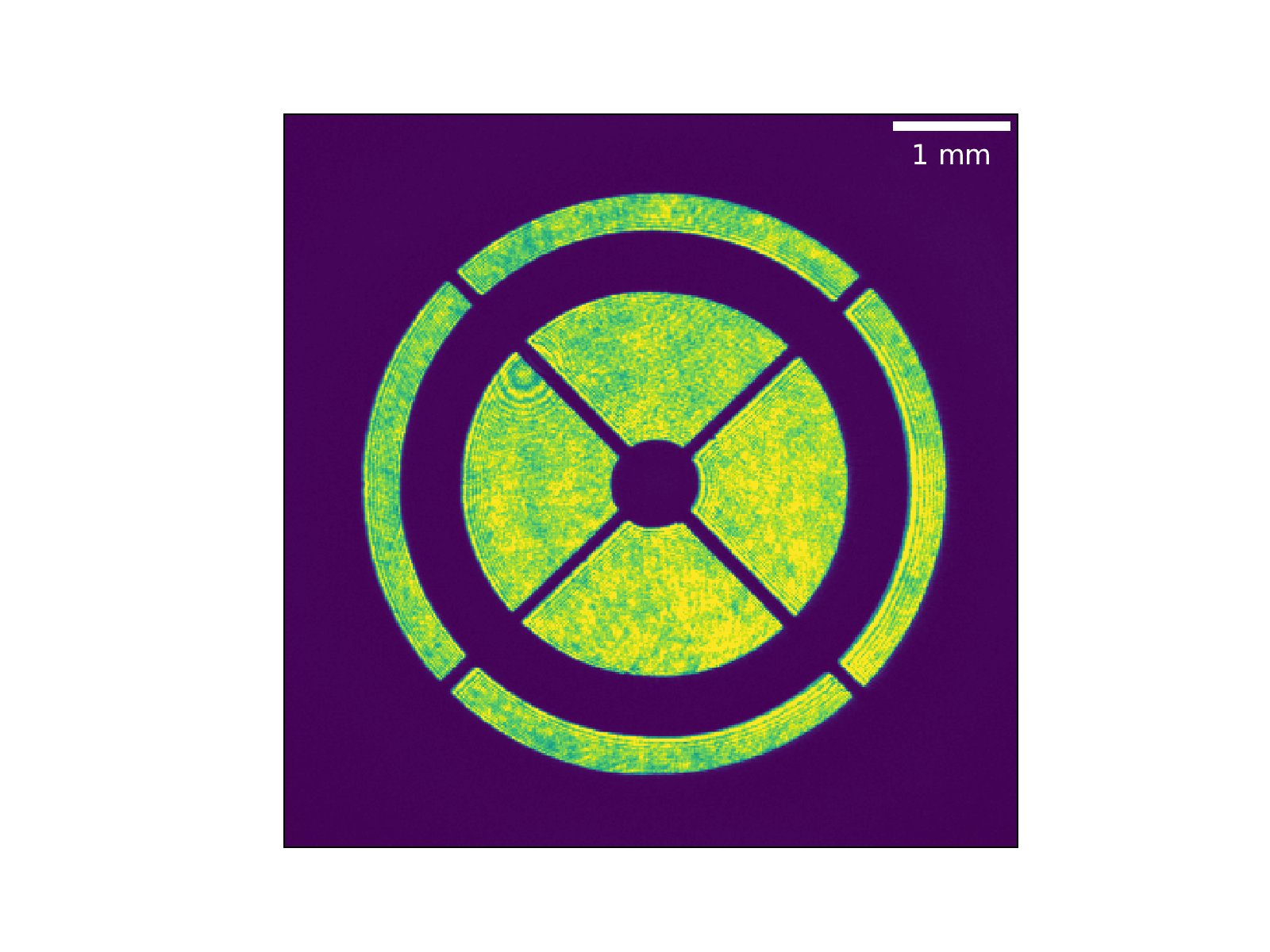} \\
\includegraphics[scale=.3]{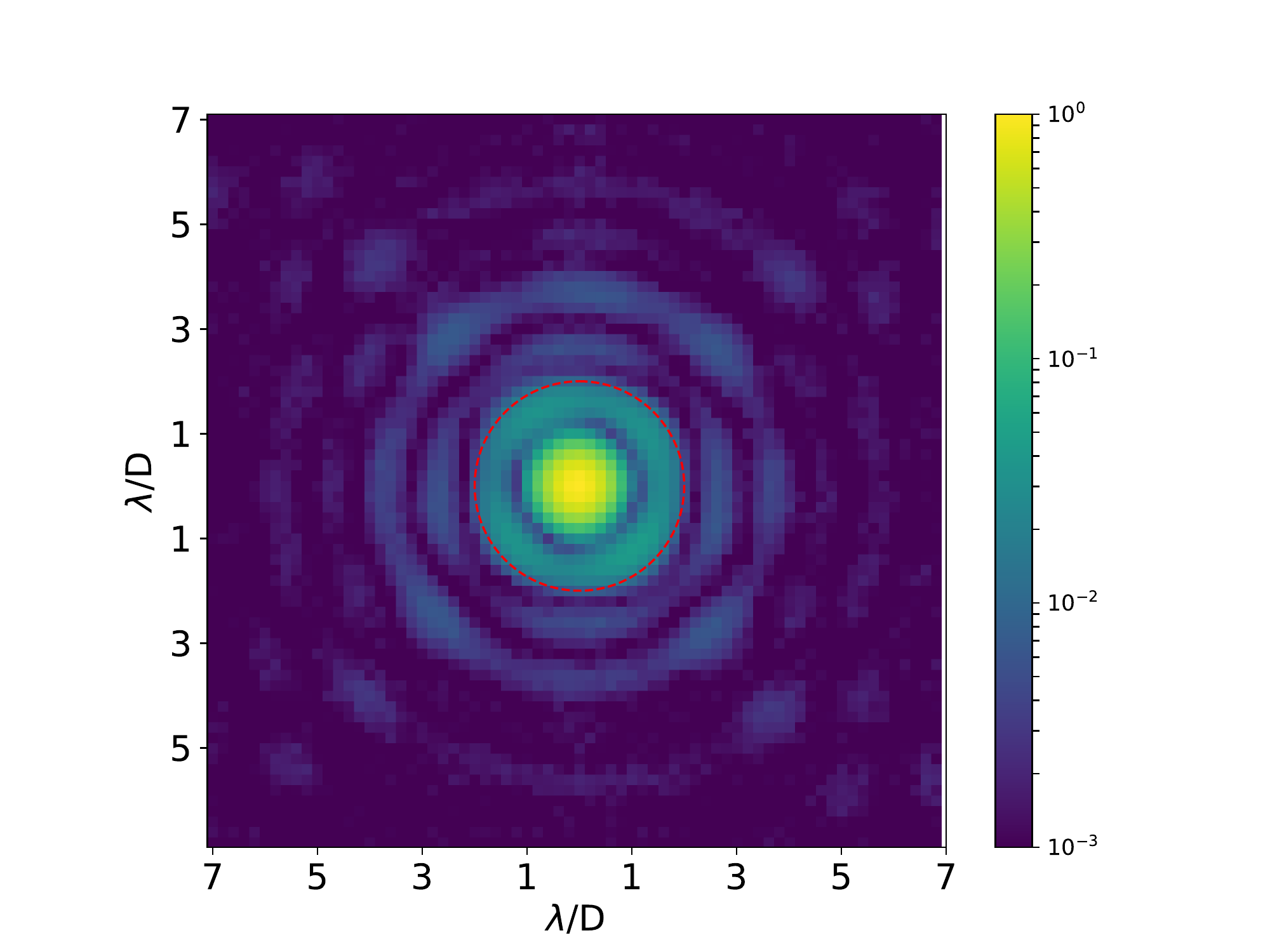} &
\includegraphics[scale=.3]{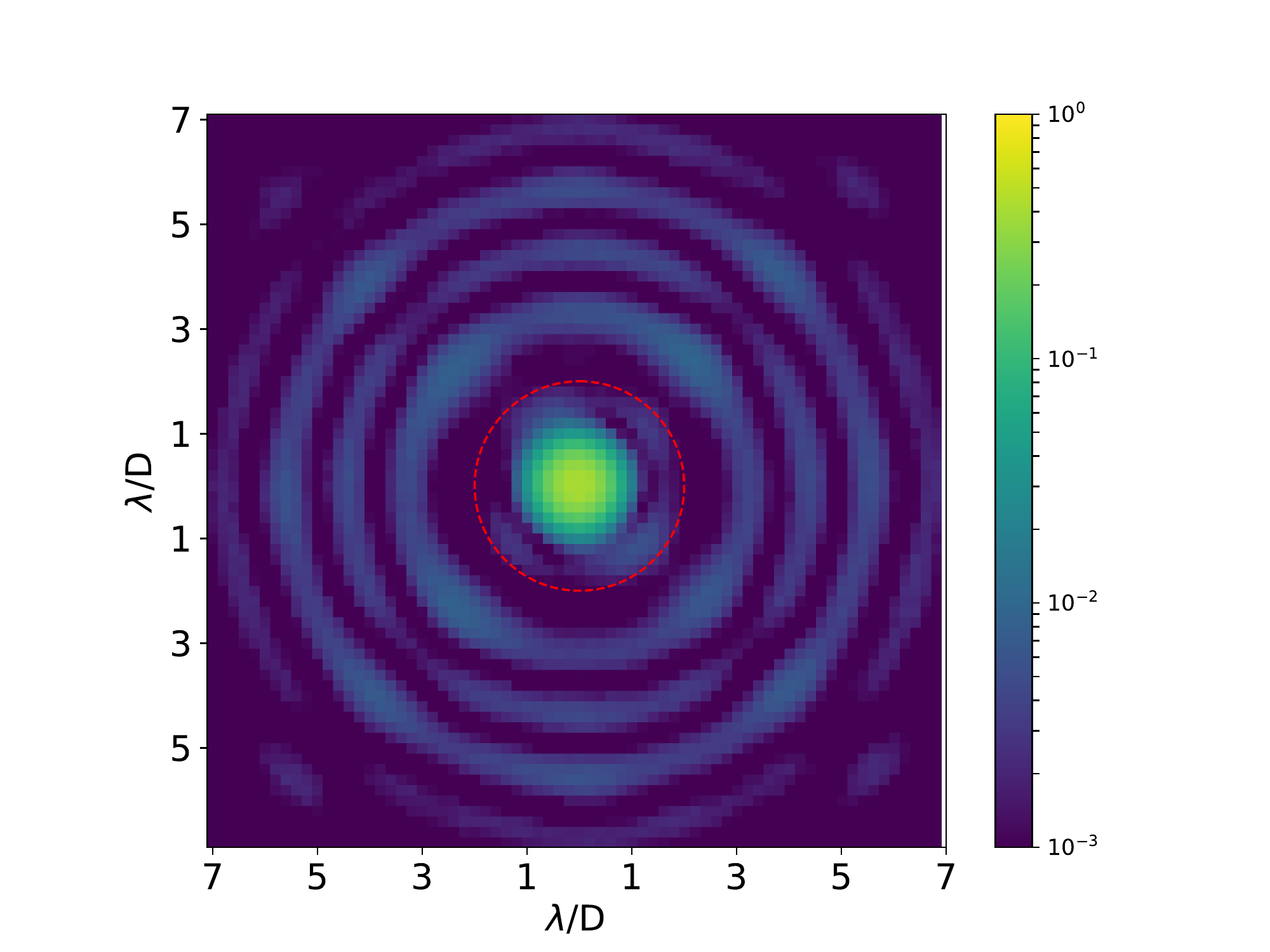}
\\
\multicolumn{2}{c}{\includegraphics[scale=.45]{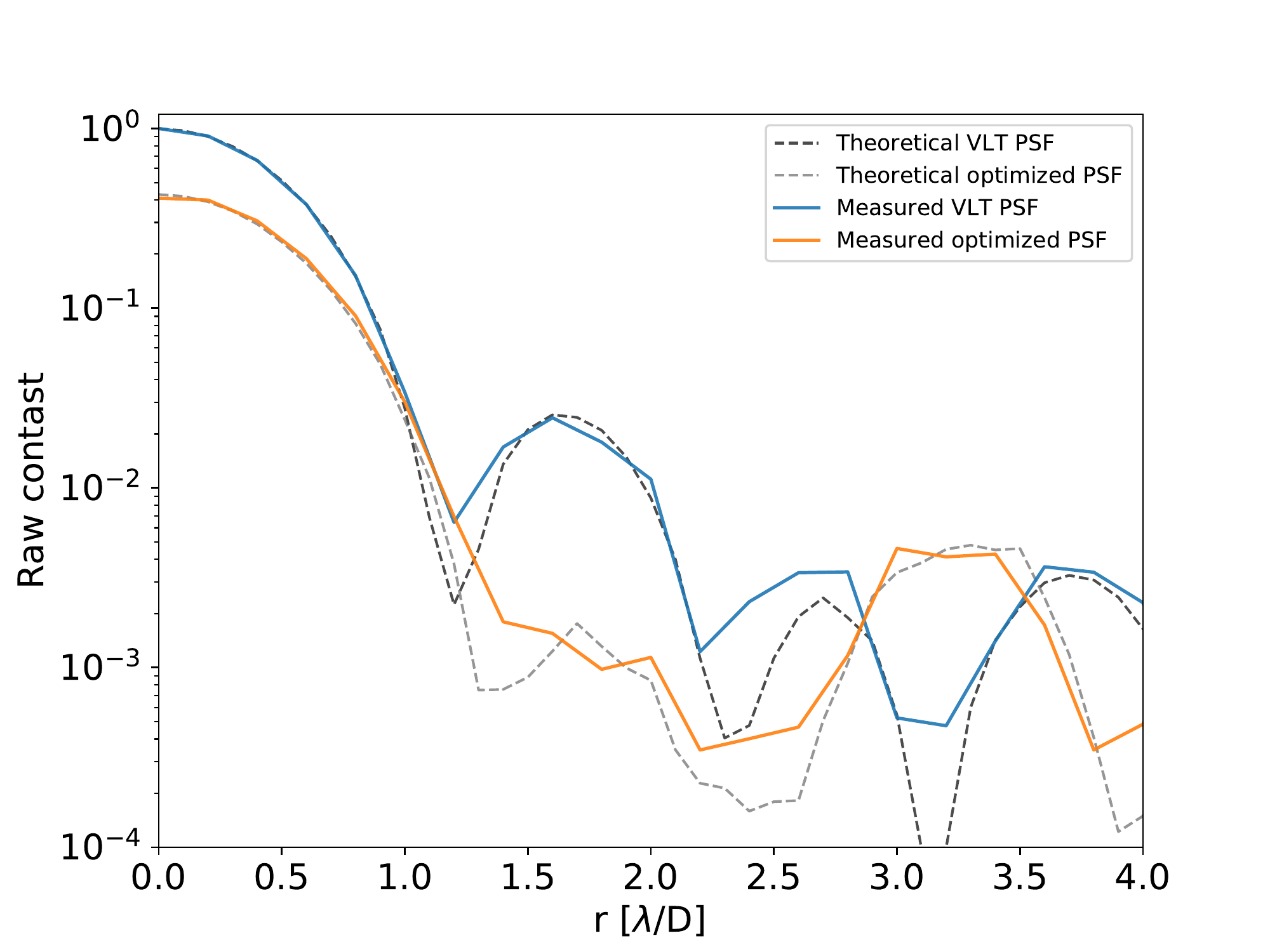}}
\end{tabular}
\end{center}
  \caption{Top: The imaged pupil plane images showing the original VLT aperture on the left and the optimized mask on the right. Middle: The measured PSF for each aperture, shown on log scale. The red dashed circle indicates the separation of 2 $\lambda$/D. Bottom: The raw contrast curve for the measured PSF is shown in color and the dashed lines indicate the raw contrast of the simulation for the same apertures.}
\label{Measured-PSF}
\end{figure}

\begin{figure}[!ht]
\begin{center}
\begin{tabular}{l l}
\includegraphics[scale=.45]{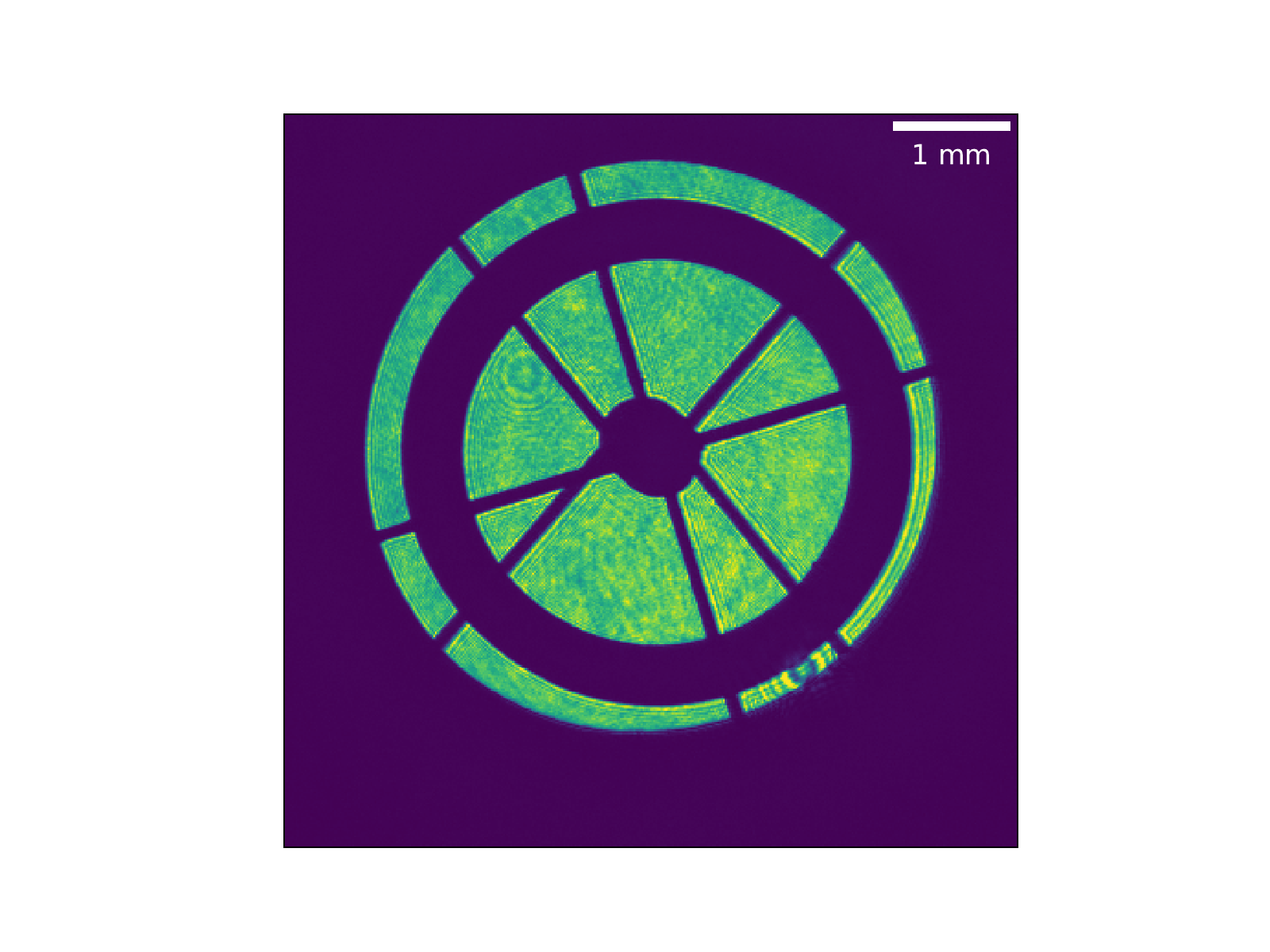}&

\includegraphics[scale=.35]{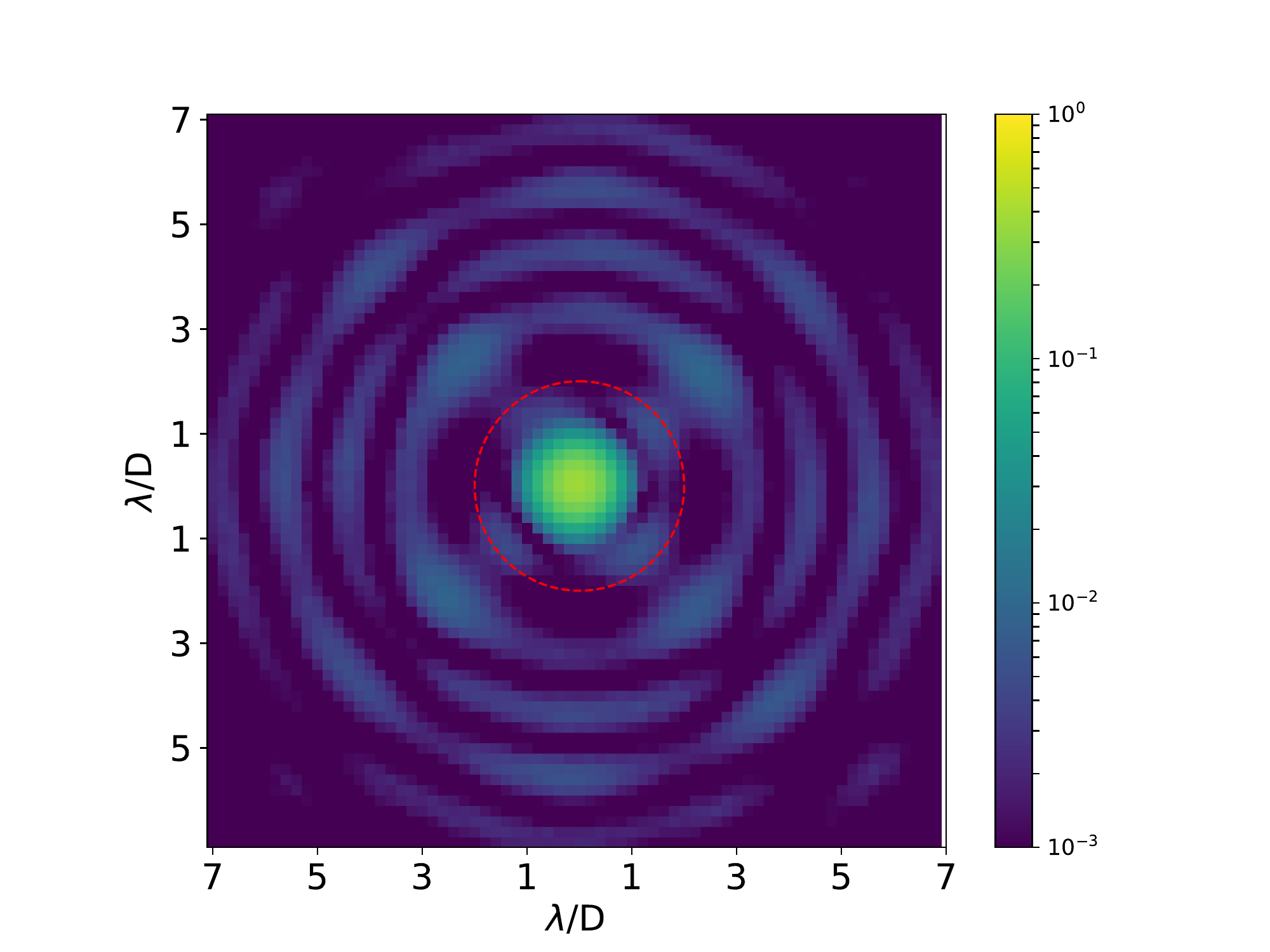} 
\\
\multicolumn{2}{c}{\includegraphics[scale=.5]{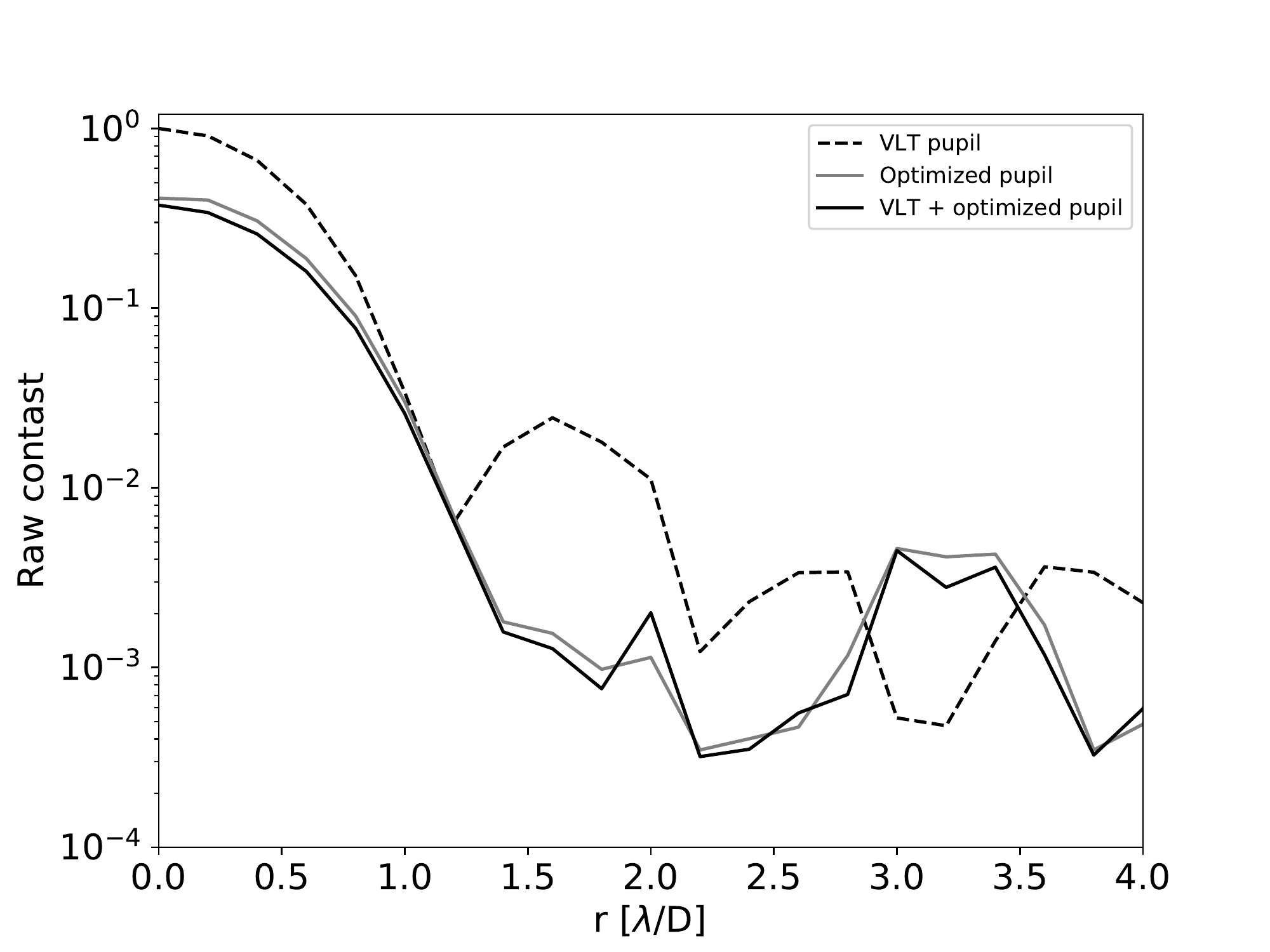}}
\end{tabular}
\end{center}
  \caption{Top: On the left both pupil masks are inserted into our test setup in successive pupil planes with some misalignment present and right the measured PSF, shown in log scale. The red dashed circle indicates the separation of 2 $\lambda$/D. Bottom: The raw contrast curve for the measured PSF as a function of separation for the different apertures.}
\label{Bothmasks}
\end{figure}

The masks were inserted into the entrance pupil and the pupil and focal plane images were recorded. The only additional step was to subtract dark frames, correct for the different exposure time of each PSF and normalize the counts with respect to the VLT pupil PSF maximum counts. The results are shown in Figure \ref{Measured-PSF} with the top panels showing the pupil images and the middle their respective PSF. We use the same logarithmic scaling as the simulation, with the minimum of the scale set to $10^{-3}$. In the bottom image of Figure \ref{Measured-PSF} the raw contrast is plotted, with color for the measured PSF and the dashed lines for the theoretical. Finally we test the two masks inserted in successive pupil planes to illustrate a more realistic scenario, where the masks are not perfectly aligned and the spiders of the optimized mask and the VLT pupil are at different positions as can be seen in Figure \ref{Bothmasks}.

\section{Discussion}

Imaging exoplanets at separations smaller than 3 $\lambda$/D and with contrasts deeper than $10^{-6}$ is extremely challenging with the current technology but the importance of the planets in this parameter space is undeniable. With Proxima b as the main target of interest and using ZIMPOL to gather polarimetric data the use of an amplitude mask that can be inserted into the setup is explored. The requirements mainly come from considering how an intervention to the setup would affect the polarization of the incoming beam and therefore we propose the use of a binary amplitude mask in the pupil plane. 

Using simulations and a GS algorithm we optimize and evaluate the performance of a the mask, showing that it would provide slightly over an order of magnitude of contrast improvement. We considered effects that will degrade the performance of the mask, simulating broadband filters to test whether using the mask the 150 nm wide I\_PRIM filter in ZIMPOL still provides a significant improvement and testing how the contrast changes with Strehl ratio. Based on the results of the simulation a bandwidth of 150 nm in the I band has little impact on the contrast (ROI: 4.9 $10^{-2}$) and with extreme AO we expect Strehl ration of 40\% to 70\% at these wavelengths. The pupil registration error also scales linearly in logarithmic space for small misalignments, which is what might be expected on the telescope.

The performance of the manufactured mask approaches the predictions in the monochromatic case. This is an encouraging sign to attempt and manufacture more complicated designs that might provide even deeper contrast. For the case of pupil misalignment the performance is only slightly affected. Eventually the contrast will degrade for realistic conditions of wavefront errors, bandwidth and misalignment of the pupil combined. Since the design of the mask and performance evaluation does not seem to be limited by manufacturing, these effects can be quantified more extensively with simulations, for example using the CAOS\cite{boccaletti2008end} simulation environment created by the SPHERE team. A summary of the results for the design of the mask that was tested is shown in Table \ref{summarytable}.

\begin{center}
\begin{table}[!ht]
\centering
\renewcommand{\arraystretch}{1}
\caption{Amplitude pupil mask characteristics for SPHERE/ZIMPOL. }
\begin{tabular}{c c c c   }
& Geometrical features & & \\ 
&   & & \\ 

 Radius of ring center  & Ring width  & Opt. Region  & Throughput\\
 (fraction of R/mm)           &  (fraction of R) & in $\lambda/D$ & \\
 \hline
 &   & & \\ 
     0.765  					   &  0.205            &      1.5-2.5          & 42\%  \\                

&   & & \\ 
 & Contrast gain in ROI & & \\ 
& & \\ 
&  & & \\ 
 Simulation  &  Measured & &\\
\hline
&   & & \\ 
11.1  &        10.6    & \\
&   & & \\

&  Mask dimensions & & \\
&   & & \\
Outer radius       &  Inner radius  &  Ring radius & Ring width \\
\hline
&   & & \\ 
5 mm			 &   0.75 mm &       3.37 mm & 0.58 mm    \\
&   & & \\ 
Spider thickness & & & \\ 
\hline 
& & & \\ 
0.01 mm & & &\\ 
\end{tabular}
\label{summarytable}
\end{table}
\end{center}

A further consideration for future work is to use better estimates of the contrast and the eventual sensitivity gain that one might get by using this mask. For this work a crude estimate was used by calculating the raw contrast, or the expected mean contrast at a given separation without considering more complex metrics. One should include also the throughput loss which could be compensated by integrating for longer. Another important feature that such an amplitude mask would provide is decreasing the gradient of the PSF in the region where the companion is located. This flattening provides a stable signal with less extreme pixel-to-pixel variation that helps when post processing the data.  

While testing further designs for creating a dark ring in the PSF, we plan to manufacture a mask with the appropriate specifications for the pupil stop wheel of ZIMPOL in order to have a mask ready in case the opportunity arises for an on-sky test. In such a case, we would first observe another target of interest, namely GJ876 b \cite{rivera20057}, an exoplanet discovered with RV located at 42 mas and demanding contrast of about $10^{-6}$ in the I band. This should be easier to detect and one would at the same time also test the performance of the mask and retrieve contrast curves after post processing. This would be informative for a future observation of Proxima.

\section{Conclusion}

The prospect of directly imaging Proxima b is exciting and we believe it is something worth pursuing even with the current technology. We show that inserting an amplitude mask in the ZIMPOL instrument would improve the contrast and provide about an order of magnitude gain in contrast. This would be a low risk high gain approach without the need of an extensive upgrade of the instrument, implementing a simple and robust concept. Even a non detection of Proxima b would be interesting, giving some limits for atmospheric models and designs of future instrument upgrades. Furthermore a detection of an exoplanet in reflected light would be of great achievement for the ZIMPOL/SPHERE instrument and pave the way for similar instruments at the future ELTs.


\acknowledgments 
We would like to thank our collaborators from the various institutes and the SPHERE consortium for their feedback and the valuable information about the SPHERE instrument. We would also like to acknowledge the mechanical workshop of the Department of Physics at ETH for their amazing work on producing these masks, and our colleagues in the Planet and Stars Formation group for their support and discussions. This work was partially funded by a PlanetS NCCR grant (TP-2018-SF6).
\bibliography{report} 
\bibliographystyle{spiebib} 

\end{document}